%% file: main_revised_2nd.tex
\documentclass[aps,prl,twocolumn,superscriptaddress]{revtex4-2}

\usepackage{graphicx}
\usepackage{dcolumn}
\usepackage{bm}
\usepackage{subfigure}
\usepackage{mathrsfs}
\usepackage{multirow}
\usepackage{amsmath}
\usepackage{orcidlink}
\usepackage{hyperref}

\usepackage{xspace}

\gdef\n4lolnl{N$^4$LO+3N$_{\rm lnl}$\xspace}
\gdef\n3lolnl{N$^3$LO+3N$_{\rm lnl}$\xspace}

\gdef\texas{N$^3$LO$_{\rm Texas}$\xspace}
\gdef\be2{$B({\rm E}2)$\xspace}

\begin{document}

\title{Neutron drip line in calcium isotopes from a chiral interaction}
\thanks{This manuscript has been authored by UT-Battelle, LLC, under contract DE-AC05-00OR22725 with the US Department of Energy (DOE). The US government retains and the publisher, by accepting the article for publication, acknowledges that the US government retains a nonexclusive, paid-up, irrevocable, worldwide license to publish or reproduce the published form of this manuscript, or allow others to do so, for US government purposes. DOE will provide public access to these results of federally sponsored research in accordance with the DOE Public Access Plan (\href{http://energy.gov/downloads/doe-public-access-plan}{http://energy.gov/downloads/doe-public-access-plan}).}

\author{B. S. Hu\orcidlink{0000-0001-8071-158X}} 
\email{baishanhu@pku.edu.cn}
\affiliation{School of Physics, and State Key Laboratory of Nuclear Physics and Technology, Peking University, Beijing 100871, China}
\affiliation{Cyclotron Institute and Department of Physics and Astronomy, Texas A\&M University, College Station, Texas 77843, USA}
\affiliation{Physics Division, Oak Ridge National Laboratory, Oak Ridge, Tennessee 37831, USA}

\author{A. Ekstr{\"o}m}
\affiliation{Department of Physics, Chalmers University of Technology, G{\"o}teborg SE-412 96, Sweden}

\author{C. Forss{\'e}n}
\affiliation{Department of Physics, Chalmers University of Technology, G{\"o}teborg SE-412 96, Sweden}

\author{G. Hagen} 
\affiliation{Physics Division, Oak Ridge National Laboratory, Oak Ridge, Tennessee 37831, USA} 
\affiliation{Department of Physics and Astronomy, University of Tennessee, Knoxville, Tennessee 37996, USA}

\author{W. G. Jiang} 
\affiliation{Mainz Institut für Theoretical Physics and PRISMA+ Cluster of Excellence, Johannes Gutenberg-Universität, Mainz 55128, Germany} 

\author{T. Miyagi} 
\affiliation{Center for Computational Sciences, University of Tsukuba, 1-1-1 Tennodai, Tsukuba 305-8577, Japan} 

\author{T. Papenbrock\orcidlink{0000-0001-8733-2849}} 
\affiliation{Department of Physics and Astronomy, University of Tennessee, Knoxville, Tennessee 37996, USA} 
\affiliation{Physics Division, Oak Ridge National Laboratory, Oak Ridge, Tennessee 37831, USA}


\begin{abstract}
Interactions derived from effective field theories of quantum chromodynamics have thus far failed to bind calcium nuclei beyond neutron number $N=40$, while nuclear density functionals typically place the neutron dripline near $^{70}$Ca, at $N=50$. We present the chiral interaction \texas, a combination of two- and three-nucleon potentials at fourth and third chiral order, respectively, with low-energy constants optimized using emulator-accelerated fits to few- and many-body data.  
This interaction accurately reproduces binding energies and charge radii of key nuclei with mass number $A=3$ to $208$, important excited states, and nuclear matter near saturation. Using  \emph{ab-initio} methods, coupled cluster and valence-space in-medium similarity renormalization group, we find that the calcium two-neutron dripline extends to $^{71}$Ca.
\end{abstract}

\maketitle

{\it Introduction.$-$}
What are the limits of nuclear binding? This question lies at the heart of our understanding of how visible matter is formed in the universe. It drives experimental and theoretical research in nuclear physics and astrophysics alike~\cite{erler2012,afanasjev2013,thielemann2017,neufcourt2020,stroberg2021}. The answer requires determining the limits of the nuclear chart, defined by the proton and neutron driplines---the boundaries beyond which an additional proton or neutron cannot be bound to the nucleus. Experimentally, the proton dripline is known up to neptunium (proton number $Z=93$)~\cite{zhang2019}, while the neutron dripline has only been mapped up to neon ($Z=10$)~\cite{ahn2019}. Very neutron-rich nuclei are difficult to synthesize in the laboratory and locating the neutron dripline for heavier elements will be extremely challenging~\cite{crawford2022,tarasov2024}. Nevertheless, knowing how many neutrons can be bound in a given isotopic chain is essential for modeling nucleosynthesis, particularly the rapid neutron-capture process that produces the heavy elements in neutron-star mergers and core-collapse supernovae~\cite{mumpower2016,roederer2023}. Reliable theoretical predictions beyond experimentally accessible regions are needed.

\begin{figure}[!t]
\setlength{\abovecaptionskip}{0pt}
\setlength{\belowcaptionskip}{0pt}
\includegraphics[scale=0.41]{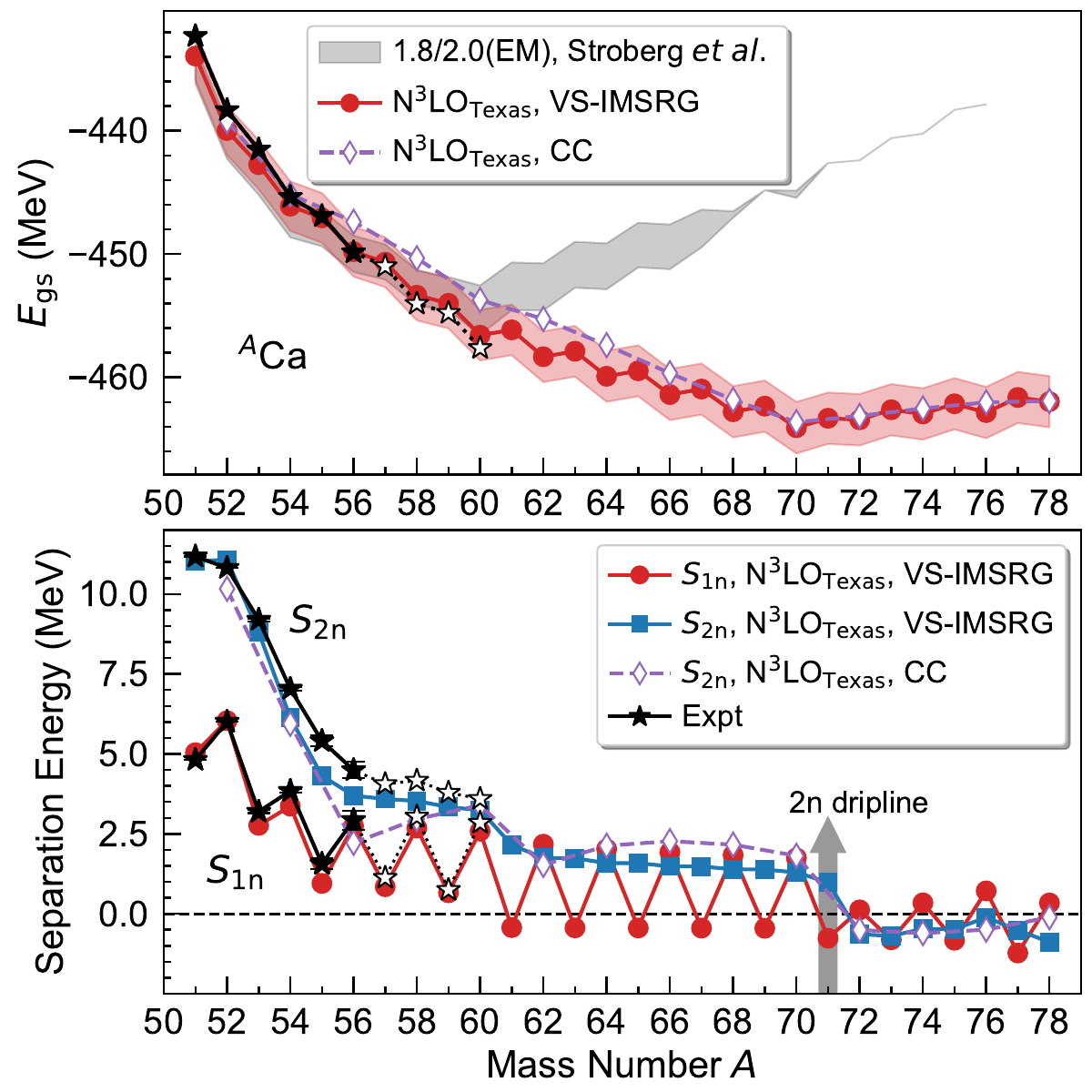} \caption{\label{Ca_BE_Sn} 
Ground-state energies ($E_{\rm gs}$), one-neutron ($S_{1n}$) and two-neutron ($S_{2n}$) separation energies for calcium isotopes computed with the interaction \texas using coupled-cluster (CC) and valence-space in-medium similarity renormalization group (VS-IMSRG) methods, compared with 1.8/2.0(EM) results~\cite{stroberg2021} and data \cite{ensdf,wang2021}. Stars (open) are masses (extrapolations) from AME2020 \cite{wang2021}. The red and gray bands indicate many-body uncertainty and different decoupled valence spaces, respectively. The CC computations are performed only for even-even nuclei.} 
\end{figure}

The calcium isotopes ($Z=20$) are especially interesting due to their closed proton shell. Energy density functional calculations predict their neutron dripline to extend to $^{70}$Ca~\cite{erler2012}, or even $^{76}$Ca~\cite{neufcourt2019}, while relativistic mean-field calculation place it between $N=42$ and 56~\cite{afanasjev2013}. Nuclear shell model calculation, based on realistic Hamiltonian derived from linked-diagram perturbation theory, shows no sign of two-neutron emission until $^{70}$Ca \cite{coraggio2020}. These predictions, which all put the dripline beyond $N=40$, are consistent with the experimental discovery of $^{60}$Ca~\cite{tarasov2018}, and measurements of the spectra of $^{56,58}$Ca and $^{62}$Ti \cite{chen2023,cortes2020}. In contrast, existing \emph{ab initio} computations based on interactions from chiral effective field theory~\cite{epelbaum2009,machleidt2011,Hammer:2019poc} place the neutron dripline closer to stability, around $^{60}$Ca~\cite{hagen2012b,hergert2014,hergert2020,li2020,stroberg2021,tichai2024}, finding that $^{61}$Ca is unbound with respect to $^{60}$Ca and further decreasing binding energy as the neutron number increases beyond $N=40$. $^{70}$Ca only appears as a probable dripline nucleus when considerable theoretical uncertainties are taken into account~\cite{stroberg2021}. Clearly, present interactions from chiral effective field theory used in \emph{ab initio} dripline predictions lack precision at best; at worst, they may exhibit a significant flaw.

In this Letter we address this shortcoming and our main results are shown in Fig.~\ref{Ca_BE_Sn}. These \emph{ab initio} results were obtained by optimizing a new chiral interaction, named \texas, and using many-body methods including coupled-cluster~\cite{kuemmel1978,bartlett2007,shavittbartlett2009,hagen2014,sun2025} and valence-space in-medium similarity renormalization group (VS-IMSRG)~\cite{tsukiyama2011,hergert2016,stroberg2017,stroberg2019}. This interaction includes higher order terms in the chiral (Weinberg) power counting compared to the previous chiral interactions \cite{ekstrom2015a,jiang2020} and has almost twice the number of low-energy constants (LECs), making the optimization much more challenging. We overcome this challenge by employing high-fidelity emulators for all observables in the objective function. We see that the predicted dripline with \texas extends up to at least $^{70}$Ca. We note that the region of neutron-rich nuclei around $N=50$ is fascinating for nuclear structure because of the interplay of weak binding, deformation, and strong inter-nucleon correlations~\cite{hamamoto2012,hagen2012b,nowacki2016,tichai2024,hu2024b}. Furthermore, one expects that subtle details of the two–nucleon (2N) and three-nucleon (3N) interactions become amplified near the dripline, and all these effects may impact its precise location~\cite{otsuka2010, kondo2023}. 

This Letter is organized as follows. We discuss the interaction, the optimization protocol, model-space parameters, and employed many-body methods leading to the results shown in Fig.~\ref{Ca_BE_Sn}. For validation we present results for the dripline in oxygen isotopes and compare with available data, we also compute binding energies and charge radii of closed-shell nuclei ranging from $^{4}$He all the way up to the heavy nucleus $^{208}$Pb, and saturation properties of infinite nuclear matter. Additional information regarding the optimization, excitation spectra in selected calcium isotopes, and the high-fidelity emulators developed in this work are presented in the End Matter and in the Supplemental Material.

{\it Methods.$-$}
We start from the intrinsic Hamiltonian
\begin{equation}
\hat{H}=\hat{T}-\hat{T}_{\rm CoM}+\hat{V}_{\rm 2N}+\hat{V}_{\rm 3N},
\label{Hint}
\end{equation}
where $\hat{T}$ and $\hat{T}_{\rm CoM}$ are the total kinetic energy and the center-of-mass kinetic energy, respectively. For the 2N potential $\hat{V}_{\rm 2N}$ and the 3N potential $\hat{V}_{\rm 3N}$, we use the chiral N$^{3}$LO$_{\rm Texas}$ potential developed in this work. 
It consists of a 2N interaction up to next-to-next-to-next-to-leading order (N$^3$LO), based on the formalism presented in Refs.~\cite{entem2015,entem2017}, and the 3N interaction up to next-to-next-to-leading order (NNLO) from~\cite{epelbaum2002}, with nonlocal regulator functions $f(p)=\exp\left[-(p/\Lambda)^{2n}\right]$ and cutoff $\Lambda$= 394 MeV with power $n=4$. Note that the 2N and 3N sector interactions come from different orders. We use a multi-start strategy to optimize the values of the 28 low-energy constants (LECs) in the 2N and 3N sectors to neutron-proton scattering phase shifts, nucleon-nucleon $^1S_0$ effective range parameters, and bound-state observables in $^2$H, $^4$He, and $^{16}$O. This latter nucleus is important because (i) it is sensitive to partial waves in the interaction beyond those probed in the lightest nuclei; (ii) we can build high-fidelity emulators for this nucleus; and (iii) experience shows that nuclear saturation is hard to reproduce without this anchor \cite{ekstrom2015a,jiang2020, maris2022, jiang2024}.
Importantly, the use of emulators reduces the computational cost of the optimization by several orders of magnitude and facilitates exploration across a broad LEC parameter domain. 
For nucleon-nucleon scattering we employed the eigenvector continuation~\cite{duguet2024} formulation presented in~\cite{melendez2021}; for $^2$H and $^4$He, we used the eigenvector continuation framework tailored to full configuration interaction (FCI) and Jacobi no-core shell model (J-NCSM)~\cite{konig2020,navratil2000} calculations, respectively; for the $^{16}$O ground-state energy and charge radius, we used the sub-space projected coupled-cluster (SPCC) method \cite{ekstrom2019}. 
In a final step of the optimization protocol we computed the binding energies of $^{22,24}$O and $^{40,48}$Ca for a few candidate interactions to identify the optimal one. We note that there was a tension between the reproduction of the scattering phase shifts in the partial-wave $^3P_2$ channel and the charge radius of $^{16}$O, consistent with the global sensitivity analysis in~\cite{ekstrom2019}.

We express the Hamiltonian given in Eq.~\ref{Hint} in a spherical harmonic oscillator basis with  spacing $\hbar\omega$ and single-particle energies up to $(e_{\rm max}+3/2)\hbar\omega$, as defined below. The 3N interaction has an additional energy cut on the allowed excitation of three nucleons given by $E_{\rm 3max}\hbar \omega$. With the Hamiltonian expressed in the harmonic oscillator basis we perform a Hartree-Fock calculation which serves as a starting point for our many-body computations. The Hamiltonian is normal-ordered with respect to the Hartree-Fock basis and the 3N interaction is approximated at the normal-ordered two-body level~\cite{hagen2007,roth2012,ripoche2020,Frosini:2021tuj,miyagi2022,hebeler2023,rothman2025}. 

In our computations of the calcium and oxygen isotopes we employ the multishell VS-IMSRG~\cite{stroberg2017,miyagi2020}, and for the calcium isotopes we also perform coupled-cluster computations that start from a reference state that breaks rotational symmetry~\cite{hagen2022,hu2024a,hu2024b,sun2025}.
Our VS-IMSRG calculations start from a model-space within $e_{\rm max}=14$, $E_{\rm 3max}= 28$, and $\hbar\omega=12$~MeV. We use the VS-IMSRG~\cite{miyagi2020} to decouple the full $A$-body Hamiltonian within a spherical Hartree-Fock basis into a multishell valence-space Hamiltonian. Specifically we employ a neutron \{$1p_{3/2},1p_{1/2},0f_{5/2},0g_{9/2},1d_{5/2},2s_{1/2}$\} valence space with $^{48}$Ca core for $^{49-78}$Ca, a neutron $p-$shell for $^{13-15}$O, and a neutron \{$sd$-shell +$0f_{7/2},1p_{3/2}$\} valence space with $^{16}$O core for $^{17-32}$O. We then perform exact diagonalizations with the KSHELL code \cite{noritaka2019}. Our coupled-cluster calculations start from a product state built from axially symmetric natural orbitals~\cite{tichai2019, novario2020, scalesi2025} computed for $e_{\rm max}$=14, $E_{\rm 3max}= 28$, and $\hbar\omega=12$~MeV. To facilitate the coupled-cluster computations, the deformed natural orbital basis are subsequently truncated to an $e_{\rm max}$=12 space following Ref.~\cite{hoppe2021}. We use the coupled-cluster singles-and-doubles approximation (CCSD), and we add 10\% of the CCSD correlation energy as an estimate of neglected triples excitations~\cite{sun2022,sun2025}, and a small correction from angular momentum projected Hartree-Fock~\cite{novario2020,hagen2022}. 

\begin{figure}
\setlength{\abovecaptionskip}{0pt}
\setlength{\belowcaptionskip}{0pt}
\includegraphics[scale=0.42]{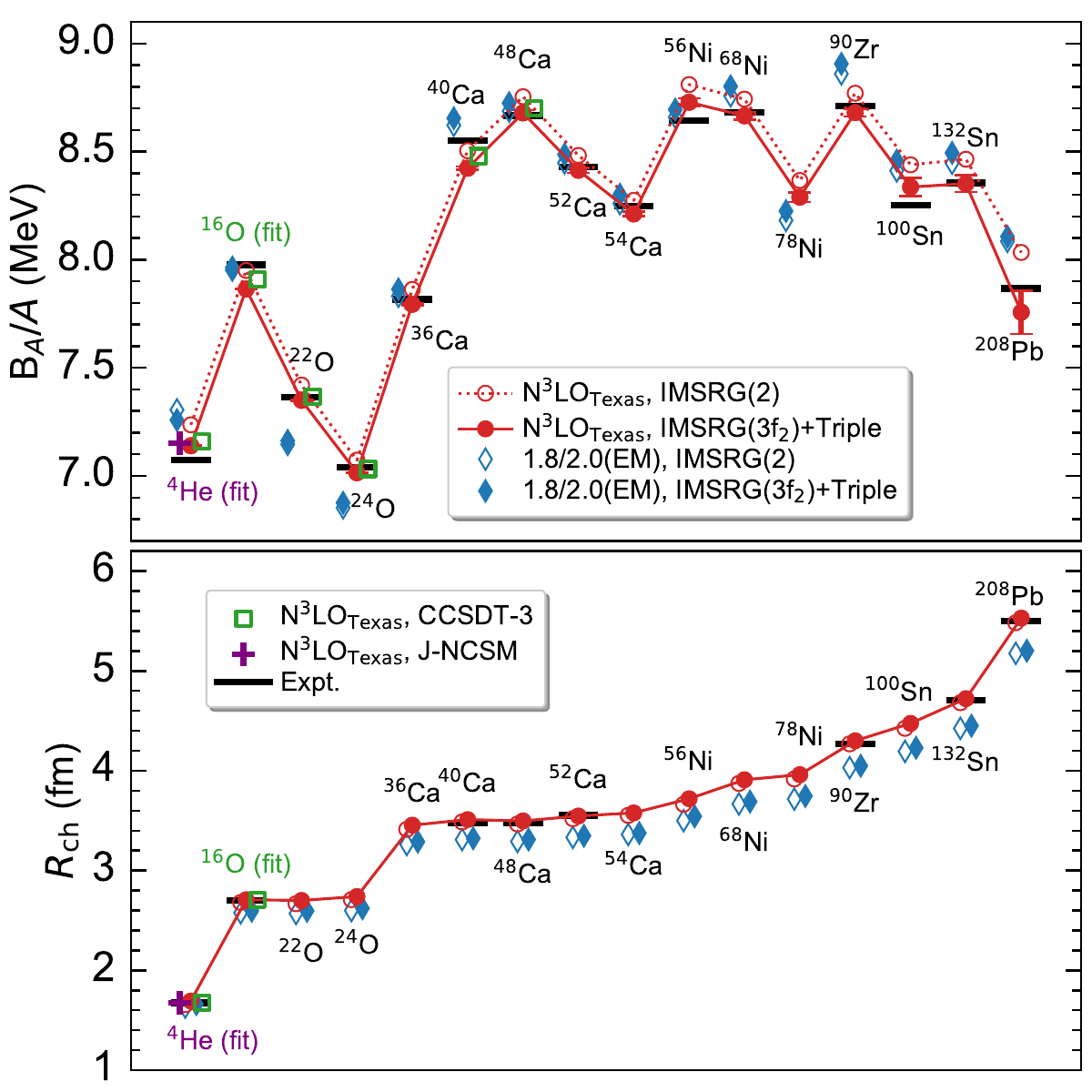} 
\caption{\label{BE_Rch} Binding energies per nucleon and charge radii for doubly closed-shell nuclei from $^{4}$He to $^{208}$Pb, computed with the J-NCSM, IMSRG(2), IMSRG(3f2) and CCSDT-3 methods using \texas, and compared to 1.8/2.0(EM) and experiment \cite{ensdf,angeli2013,garciaruiz2016}. Error bars on the IMSRG(3f2) results represent the uncertainty from extrapolation to the infinite model space. In the legend, ``Triple" indicates the inclusion of perturbative triples \cite{he2024}. 
}
\end{figure}

For closed-shell nuclei, we use IMSRG(2) and the approximate IMSRG(3)~\cite{heinz2021,stroberg2024} scheme known as IMSRG(3f2)~\cite{he2024}, as well as coupled-cluster computations with leading-order iterative triples corrections, CCSDT-3~\cite{noga1987}. For the energy computations shown in the upper panel of Fig.~\ref{BE_Rch} we performed extrapolations to the infinite basis limit following Ref.~\cite{hu2022}.
Energies computed by IMSRG(2) and IMSRG(3f2) were extrapolated using $e_{\rm max}$=8, 10, 12, 14 with $\hbar\omega=16$~MeV (except for $^{208}$Pb, where we used $\hbar\omega=12$~MeV) and fixed $E_{3\rm max}$=28. For $^{208}$Pb, we additionally included extrapolated energies based on $E_{3\rm max}$=18, 20, 22, 24, 26 and 28, as well as the difference between $e_{\rm max}$=14 and $e_{\rm max}$=12 calculations. CCSDT-3 results of $^{40}$Ca and $^{48}$Ca were extrapolated from $e_{\rm max}$=8, 10, 12 with fixed $E_{3\rm max}$=16, and those for $^{4}$He and $^{16,22,24}$O were extrapolated from $e_{\rm max}$=8, 10, 12, 14 with the same $E_{3\rm max}$=16. All VS-IMSRG charge radii were calculated using $e_{\rm max}$=14 and $E_{3\rm max}$=28, and CCSDT-3 used $e_{\rm max}$=12 and $E_{3\rm max}$=16. We have verified that $E_{3\rm max}$=16 is sufficient to achieve convergence for light and medium-mass nuclei. For infinite nuclear matter, we use periodic boundary conditions and perform coupled-cluster computations with the Taube–Bartlett $\Lambda$-CCSD(T) approximation~\cite{taube2008}, on a discrete lattice in momentum space~\cite{hagen2014b,marino2024}. For few-nucleon systems ($A\leq4$), we employed the J-NCSM to obtain virtually exact results~\cite{navratil2000}.

{\it Results.$-$}
Figure~\ref{Ca_BE_Sn} shows the calculated ground-state energy ($E_{\rm gs}$), and separation energies ($S_{1n}$ and $S_{2n}$) of neutron-rich calcium isotopes, compared with the 1.8/2.0(EM) results from \citet{stroberg2021}. In our coupled-cluster calculations,  $^{56}$Ca and $^{58}$Ca are nearly spherical, and we do not include the estimated energy contribution from angular momentum projection. This omission is one reason for the discrepancy observed between the coupled cluster and VS-IMSRG results. 
The predicted $S_{1n}$ and $S_{2n}$ values for calcium isotopes agree with experiment~\cite{ensdf}. The \texas interaction predicts that $^{61}$Ca is unbound with respect to $1n$ emission, while the $2n$ dripline extends to $^{71}$Ca. These results are consistent with the Bayesian posterior probabilities for neutron binding in a Bayesian model average of global density functional theory models~\cite{neufcourt2019}. 
In comparison, the energies computed by~\citet{stroberg2021} place the dripline around $^{60}$Ca. However, when those results are combined with experimental data in a Bayesian linear regression model then there is an approximate 50\% probability that calcium isotopes heavier than $^{60}$Ca are bound~\cite{stroberg2021}. With the \texas interaction we found that the $N=40$ sub-shell gap in heavy-mass calcium nuclei is smaller, and this pushes the $2n$ dripline toward $^{70}$Ca. This is consistent with expectations from the discovery of $^{60}$Ca \cite{tarasov2018} and measured spectra of $^{56,58}$Ca and $^{62}$Ti \cite{chen2023,cortes2020}. The computed spectra of $^{50,52,54,56,58}$Ca agree with large-scale shell model calculations~\cite{honma_2005} and available experimental data. The positive parity states are of similar quality as that obtained with the 1.8/2.0(EM) interaction, whereas the negative-parity states lie significantly lower, reflecting the smaller $N=40$ gap (see Supplemental Material). 

We now turn to neutron-rich oxygen isotopes as they provide a stringent test for chiral interactions and quantum many-body methods~\cite{kondo2023}. The neutron dripline in oxygen isotopes is at $^{24}$O~\cite{hoffman2009,kanungo2009,otsuka2010}, with the ground states of $^{25}$O and $^{26}$O unbound by approximately 830~keV \cite{jones2017} and 18~keV~\cite{lunderberg2012,kondo2016}, respectively, relative to the $^{24}$O ground state. Recently, \textcite{kondo2023} observed that both $^{27}$O and $^{28}$O exist as narrow unbound resonances.

\begin{figure}[!t]
\setlength{\abovecaptionskip}{0pt}
\setlength{\belowcaptionskip}{0pt}
\includegraphics[scale=0.42]{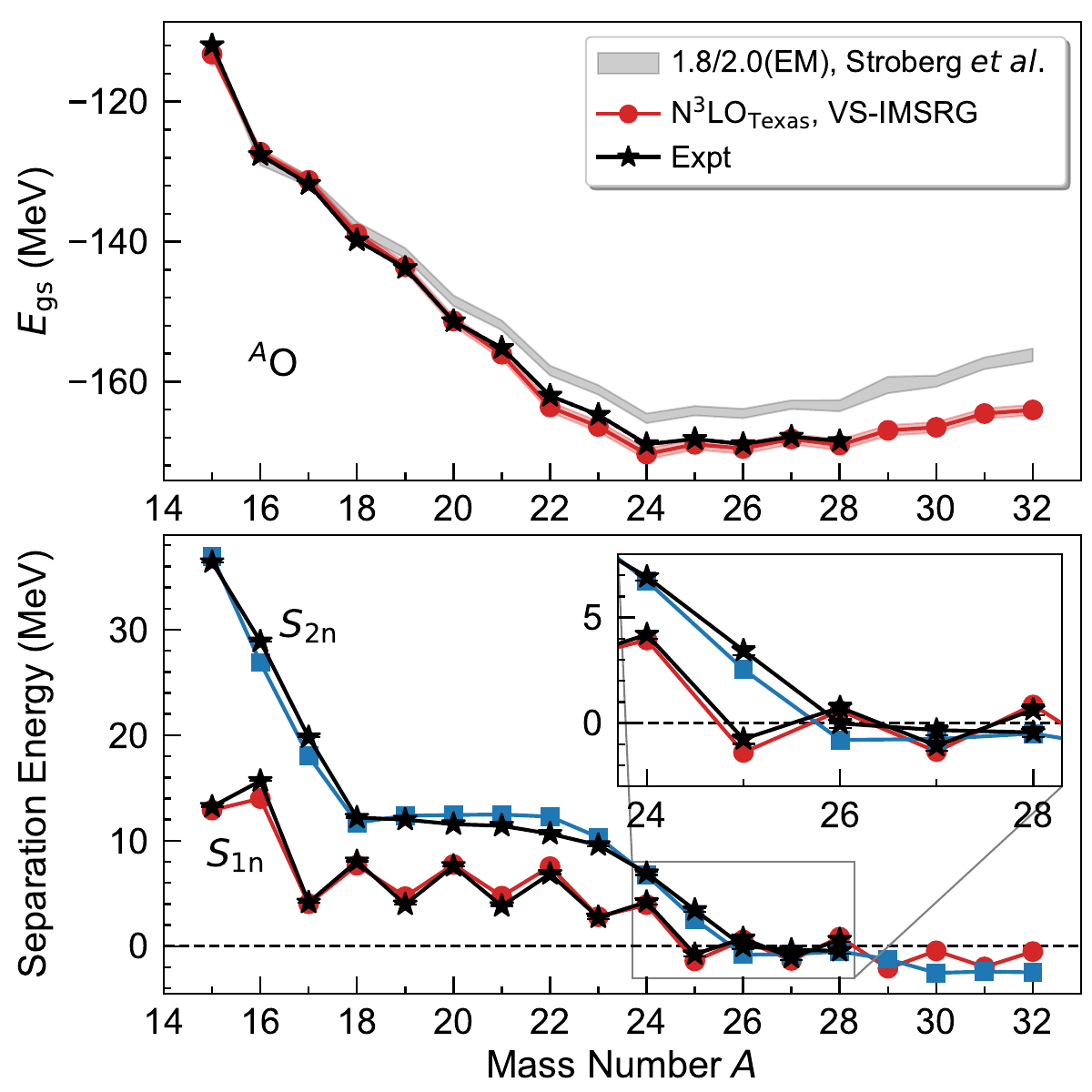} \caption{\label{O_BE_Sn}
Similar to Fig.~\ref{Ca_BE_Sn} but for oxygen isotopes.
Experimental data taken from Refs.~\cite{ensdf,kondo2023}.
}
\end{figure}

Figure~\ref{O_BE_Sn} compares the VS-IMSRG results with the N$^3$LO$_{\rm Texas}$ interaction to data and to those from \textcite{stroberg2021}. 
The top panel shows that ground-state energies are accurately reproduced while the bottom panel shows accurate results for the $S_{1n}$ and $S_{2n}$. The thin red band for VS-IMSRG with \texas indicates the estimated many-body uncertainty, quantified as the $E_{\rm gs}$ difference between ``IMSRG(3f2)+Triple" and ``IMSRG(2)" for oxygen and calcium isotopes listed in Fig.~\ref{BE_Rch}, which is approximately 1\%.  
The gray band for the 1.8/2.0(EM) shows the uncertainty arising from different decoupled valence spaces. We note that the results by \textcite{stroberg2021} were based on $\hbar\omega=16$~MeV, whereas we use a smaller  $\hbar\omega=12$~MeV for both oxygen and calcium isotopes to better capture weak binding effects near the dripline region. We checked that our dripline predictions are independent of $\hbar\omega$, and we also found accurate results for fluorine isotopes (see Supplemental Material).

The \texas interaction also reproduces low-lying spectra like the first $2^+$ excitation energies in doubly-magic oxygen isotopes, computed using the equation-of-motion method EOM-CCSDT-3 with $e_{\rm max}$=~8 and 10, see Tab.~\ref{tab:observables} in the End Matter.

\begin{figure}
\setlength{\abovecaptionskip}{0pt}
\setlength{\belowcaptionskip}{0pt}
\includegraphics[scale=0.42]{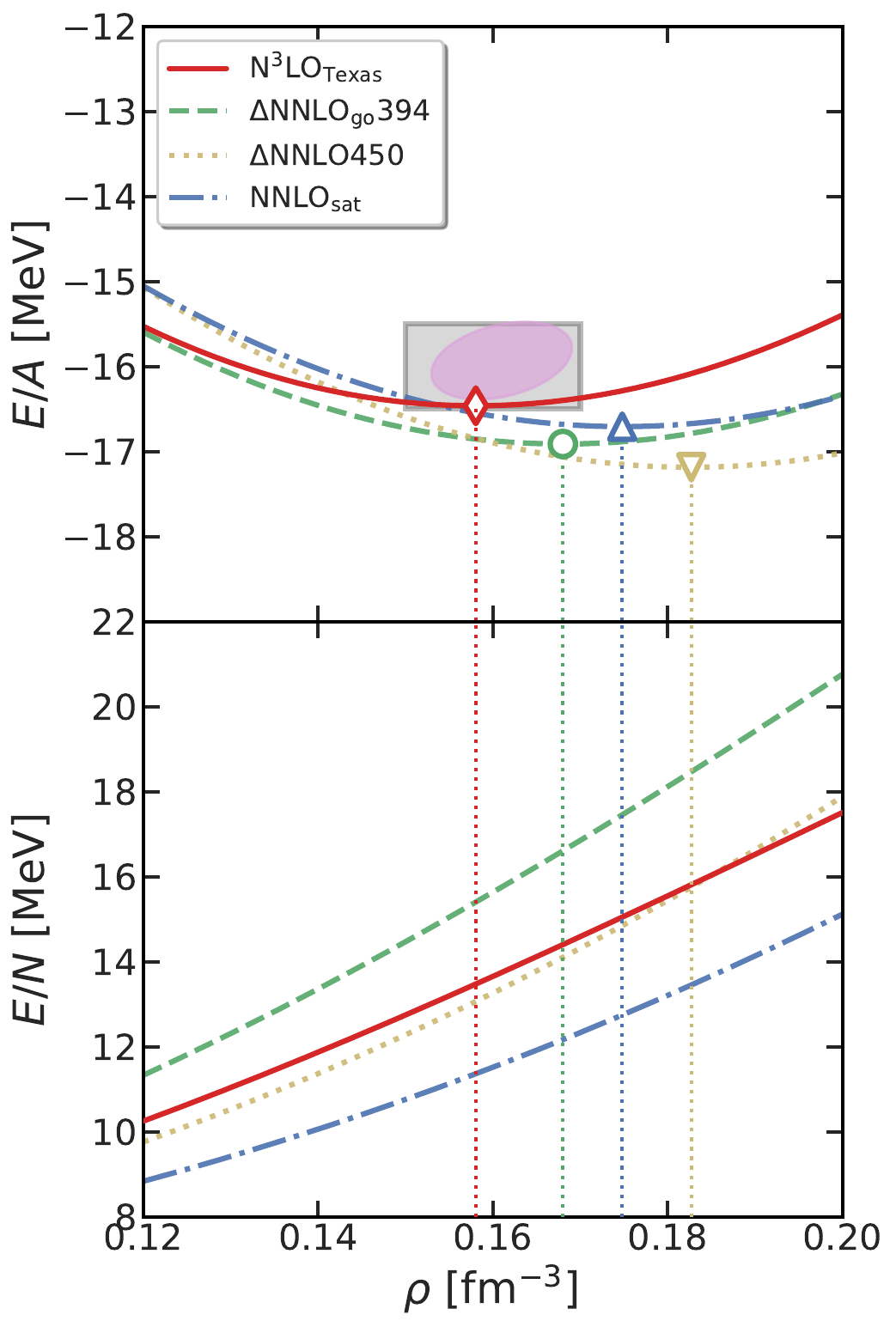} 
\caption{\label{NM} Energy per nucleon for symmetric nuclear matter (top) and pure neutron matter (bottom). The gray rectangle marks the empirical saturation region
from Ref.~\cite{drischler2021}. The pink ellipse shows a 95\% confidence region for the saturation point inferred from Bayesian posterior distributions based on a set of density functional theory predictions \cite{drischler2024}.
}
\end{figure}

Finally we turn to bulk properties of key closed-shell nuclei ranging from $^4$He to $^{208}$Pb and saturation properties of infinite nuclear matter. Figure~\ref{BE_Rch} presents the calculated binding energies and charge radii using the VS-IMSRG and coupled-cluster methods and compares them with results from the 1.8/2.0(EM) interaction. The \texas interaction accurately describes these bulk properties up to $^{208}$Pb with deviations below $1.5\%$. We note that the $N=Z$ doubly-magic nuclei $^{40}$Ca, $^{56}$Ni, and $^{100}$Sn deviate a bit more from data than the other nuclei. This could be because $^{16}$O was optimized using the computationally expensive CCSDT-3 method, while the heavier nuclei were computed using the more affordable IMSRG(3f2) method.

Figure~\ref{NM} shows our predictions for the equation of state of nuclear matter, compared with the results from the $\Delta$NNLO$_{\rm GO}$(394)~\cite{jiang2020,marino2024}, $\Delta$NNLO(450)~\cite{ekstrom2018,marino2024}, and NNLO$_{\rm sat}$~\cite{ekstrom2015a} interactions. Although nuclear-matter properties were not included in the optimization, the N$^3$LO$_{\rm Texas}$ prediction is located within the empirically constrained regions of nuclear saturation~\cite{drischler2021,drischler2024}. 
In this work we employed the $\Lambda$-CCD(T) method~\cite{taube2008}, which includes more accurate triples corrections compared to the CCD(T) approximation~\cite{raghavachari1989} used in Ref.~\cite{marino2024} (see Supplemental Material).

{\it Summary and outlook.$-$}
By developing a novel chiral interaction, \texas, we resolve the long-standing tension over the calcium neutron dripline, reconciling \emph{ab initio} predictions with density functional theory and existing experimental results. This interaction also reproduces nucleon-nucleon scattering phase shifts, charge radii, spectra, and nuclear-matter properties. The key ingredients were (i) employing a regulator-consistent chiral 2N potential at N$^3$LO and a 3N potential at NNLO, (ii) jointly calibrating 2N and 3N LECs to selected few-nucleon data and the binding energy and charge radius of $^{16}$O, while validating against heavier-mass calcium and neutron-rich oxygen isotopes, and (iii) using emulator-accelerated multi-start optimization algorithms to navigate the high-dimensional parameter space spanned by the LECs. Our results suggest that \texas yields energy results that are comparable to the chiral interaction 1.8/2.0(EM), while improving radii systematics and dripline physics. Notably, \texas predicts that the two-neutron calcium dripline is located at $^{71}$Ca, a significant extension from previous limits set by \emph{ab initio} calculations based on chiral interactions, but in line with recent experimental trends. A next step is to use \texas as the starting point of a Bayesian analysis that accounts for effective field theory and \emph{ab initio} many-body uncertainties \cite{wesolowski2021,jiang2024,jiang2024b,curry2025qmc}, thereby enabling posterior predictive distributions and a systematic assessment of the predictive power of consistently regulated higher-order 3N contributions~\cite{Epelbaum:2019kcf,Krebs:2023ljo,Krebs:2023gge}. The 2N and 3N matrix elements for the chiral \texas interaction are available upon request and are included in the {\tt NuHamil} code~\cite{miyagi2023}, as well as provided in Ref.~\cite{Hu2025dataset}.

\begin{acknowledgments}
We thank Ragnar Stroberg and Bingcheng He for the {\tt imsrg++} code~\cite{stroberg_imsrg} used to perform the IMSRG(2) and IMSRG(3f2) calculations. This work was supported by the U.S. Department of Energy, Office of
Science, under SciDAC-5 (NUCLEI collaboration), under grants Nos.~DE-FG02-97ER41014, DE-FG02-96ER40963, DE-SC0024465, and DE-SC0026198 (STREAMLINE collaboration),
by the Swedish Research Council (Grants No. 2020-05127, No. 2021-04507, and No. 2024-04681), 
by the Alexander von Humboldt Foundation, 
by the Cyclotron Institute at Texas A\&M University,
by JST ERATO Grant No. JPMJER2304, Japan, and by JSPS KAKENHI Grant Numbers 25K07294, 25K00995, and 25K07330.
This research used resources from the Oak Ridge Leadership Computing Facility located at Oak Ridge National Laboratory, which is supported by the Office of Science of the U.S. Department of Energy under contract No. DE-AC05-00OR22725. Advanced computing resources are also provided by Texas A\&M High Performance Research Computing.
\end{acknowledgments}

\bibliography{Ref,master}
\clearpage

\section{End Matter}
\begin{figure*}
\setlength{\abovecaptionskip}{0pt}
\setlength{\belowcaptionskip}{0pt}
\includegraphics[scale=0.42]{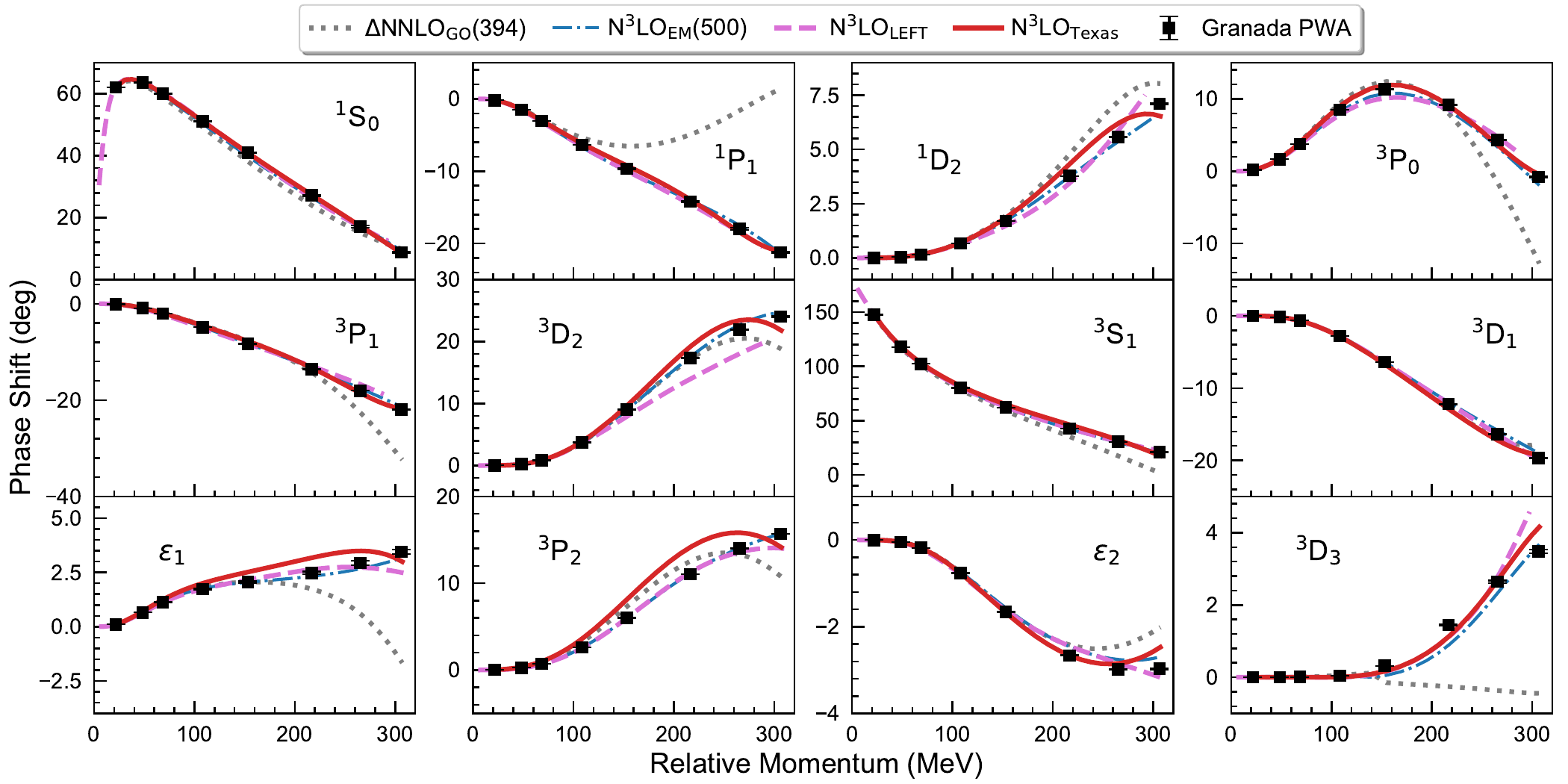} \caption{\label{phases} Neutron-proton scattering phase shifts and mixing angles as functions of relative momenta, compared with the Granada phase shift analysis \cite{perez2013} and results from the 1.8/2.0(EM)~\cite{hebeler2011,simonis2017}, $\Delta$NNLO$_{\rm GO}$(394)~\cite{jiang2020} and N$^3$LO$_{\rm LEFT}$~\cite{elhatisari2024} interactions.
}
\end{figure*}

\setlength{\tabcolsep}{4pt}
\renewcommand{\arraystretch}{1.5} 
\begin{table}[!t]
\centering 
\caption{$^1S_0$ effective range parameters $(a,r)$ in fm, ground-state energies ($E_{\rm gs}$) in MeV, charge radii ($R_{\rm ch}$) in fm, and quadrupole moment ($Q$) in fm$^2$ used to optimized the LECs. Results based on J-NCSM for $A=2,3,4$, EOM-CCSDT-3 for $^{22,24}$O, and IMSRG(3f2) for $^{78}$Ni, $^{132}$Sn and $^{208}$Pb. Results from the validation step are computed with CCSDT-3. The uncertainties in the EOM-CCSDT-3 computations stem from the truncated model space. Experiment  from~\cite{bergervoet1988,miller1990,gonzalez2006,chen2008ann,entem2017,ensdf,wang2021,angeli2013,bergervoet1988}.}
\label{tab:observables}
\begin{tabular}{llcc}
\hline\hline
 &  
 & N$^3$LO$_{\mathrm{Texas}}$
 & Exp. \\ 
\hline
\multirow{13}{*}{\rotatebox{90}{\text{Optimization}}}
& $a_{pp}^{C}$ & $-7.820$ & $-7.8196(26)$$^{a}$\\
& $r_{pp}^{C}$ & $2.757$ & $2.790(14)$$^{a}$\\
& $a_{nn}$   & $-18.950$ & $-18.95(40)$ \\
& $r_{nn}$   & $2.794$ & $2.75(11)$ \\
& $a_{np}$   & $-23.741$ & $-23.740(20)$ \\
& $r_{np}$   & $2.681$ & $2.77(5)$ \\
& $E_{\rm gs}(^{2}\mathrm{H})$ & $-2.225$ & $-2.224575(9)$ \\
& $R_{\mathrm{ch}}(^{2}\mathrm{H})$ & $2.131$ & $2.1421(88)$ \\
& $Q(^{2}\mathrm{H})$      & $0.27$ & $0.27$$^{b}$ \\
& $P_D(^{2}\mathrm{H})$      & $2.68$\% & -- \\
& $E_{\rm gs}(^{4}\mathrm{He})$              & $-28.61$ & $-28.2957$ \\
& $R_{\mathrm{ch}}(^{4}\mathrm{He})$ & $1.675$ & $1.6775(28)$ \\
& $E_{\rm gs}(^{16}\mathrm{O})$              & $-126.54$ & $-127.6193$ \\
& $R_{\mathrm{ch}}(^{16}\mathrm{O})$ & $2.707$ & $2.6991(52)$ \\
\hline
\multirow{4}{*}{\rotatebox{90}{\text{Validation}}}
& $E_{\rm gs}(^{22}\mathrm{O})$              & $-162.06$ & $-162.0272(572)$ \\
& $E_{\rm gs}(^{24}\mathrm{O})$              & $-168.77$ & $-168.9525(1680)$ \\
& $E_{\rm gs}(^{40}\mathrm{Ca})$              & $-339.15$ & $-342.0522$ \\
& $E_{\rm gs}(^{48}\mathrm{Ca})$              & $-417.63$ & $-416.0012$ \\
\hline
\multirow{10}{*}{\rotatebox{90}{\text{Prediction}}}
& $E_{\rm gs}(^{3}\mathrm{H})$              & $-8.50$ & $-8.4818$ \\
& $R_{\mathrm{ch}}(^{3}\mathrm{H})$ & $1.769$ & $1.7591(363)$ \\
& $E_{\rm gs}(^{3}\mathrm{He})$              & $-7.76$ & $-7.7180$ \\
& $R_{\mathrm{ch}}(^{3}\mathrm{He})$ & $1.967$ & $1.9661(30)$ \\
& $E_{2^+}(^{22}\mathrm{O})$ & $3.28(7)$ & $3.199$ \\
& $E_{2^+}(^{24}\mathrm{O})$ & $4.05(11)$ & $4.760$ \\
& $E_{\rm gs}(^{78}\mathrm{Ni})$              & $-647(2)$ & $-642.5640(3900)$$^{c}$ \\
& $R_{\mathrm{ch}}(^{78}\mathrm{Ni})$              & $3.961$ & -- \\
& $E_{\rm gs}(^{132}\mathrm{Sn})$              & $-1102(5)$ & $-1102.8432(19)$ \\
& $R_{\mathrm{ch}}(^{132}\mathrm{Sn})$              & $4.723$ & $4.7093(76)$ \\
& $E_{\rm gs}(^{208}\mathrm{Pb})$              & $-1613(21)$ & $-1636.4302(12)$ \\
& $R_{\mathrm{ch}}(^{208}\mathrm{Pb})$              & $5.530$ & $5.5012(13)$ \\
\hline\hline
\multicolumn{4}{l}{$^{a}$Including Coulomb (C) effects. Tuned separately after optimization.} \\
\multicolumn{4}{l}{$^{b}$CD-Bonn value~\cite{machleidt2001}} \\
\multicolumn{4}{l}{$^{c}$Trends from mass surface~\cite{wang2021}} \\
\end{tabular}
\end{table}

The \texas interaction comprises a chiral 2N potential up to N$^3$LO and a 3N potential up to NNLO. For the 2N interactions we follow the formalism presented in Refs.~\cite{entem2015,entem2017}. In this approach, the two-pion exchange bubble diagram proportional to $c^2_i$ that appears at N$^3$LO is very attractive, while the N$^4$LO two-pion exchange correction proportional to $c_i/M_N$, is large and repulsive. Therefore, as in~\cite{entem2015,entem2017}, we promote these corrections to N$^3$LO and arrive at a more realistic intermediate attraction at this order. For the subleading two-pion exchange we also use spectral function regularization~\cite{epelbaum2006} with a cutoff 700 MeV. We employ a standard nonlocal regulator function $f(p)=\exp(\left(-(p/\Lambda)^{2n}\right)$ with cutoff $\Lambda$= 394 MeV and power $n$=4 consistently for both the 2N and 3N potentials. This low-momentum cutoff provides favorable convergence with respect to model space size, enabling many-body calculations of heavy-mass nuclei without further softening the interaction via similarity renormalization group methods. The momentum cutoff and regulator power $n$ are the same as those used for the 3N potential in the 1.8/2.0(EM) interaction. However, unlike the 2N potential in the 1.8/2.0(EM), we employ $n=4$ consistently across all partial waves. 

For the pion-nucleon LECs we employ the central values obtained in the Roy-Steiner sub-threshold analysis of pion-nucleon scattering ~\cite{hoferichter2015,hoferichter2016}, see also Tab.~\ref{tab:lecs}. The statistical uncertainties of these LEC values are exceedingly small, typically at the percent level or below, with the exception of $\bar{d}_5$, which is also the numerically smallest LEC. We therefore treat the pion-nucleon LECs as fixed in the optimization of the contact LECs in the 2N and 3N potentials.

The 2N contact potential at N$^3$LO is parameterized by 26 LECs, and the 3N contacts at NNLO are parameterized by the LECs c$_{\rm D}$ and c$_{\rm E}$. In this work, we simultaneously optimize the values of these 28 LECs using the following data: low-energy neutron-proton scattering phase shifts from the Granada analysis up to 200 MeV scattering energy in the laboratory system \cite{perez2013}; the $^1S_0$ scattering lengths $a$ and effective ranges $r$ in the neutron-neutron ($nn$), neutron-proton ($np$) and proton-proton ($pp$) sectors; the deuteron quadrupole moment $Q(^2{\rm H})$; the ground-state energies ($E_{\rm gs}$) and charge radii ($R_{\rm ch}$) of $^2$H, $^4$He, and $^{16}$O. We employed reduced-basis methods~\cite{duguet2024} to develop high-fidelity emulators for all these observables, see the Supplemental Material for further details. For the bound-state observables we use an L1 objective function, while for the scattering phase shifts we use an L2 objective function. The former is more tolerant to outliers. We assign relative weights to the calibration data in the objective function to balance impact of the different kinds of data and to ensure that the calibration is not dominated by the nucleon–nucleon scattering phase shifts. These weights were selected as follows: Weight 50 for the deuteron energy, radius, and quadrupole moment; 
15 (45) for the $^4$He energy (radius); 10 (40) for the $^{16}$O energy (radius); 
60 for the $nn$ and $np$ effective range parameters; and 40 for the $pp$ parameters. For the phase-shift data, we use weights of 100 for the ${}^3S_1$ and ${}^3D_1$ channels, 100/140 for $\epsilon_1$; 9 and 0.2 for the ${}^3P_2$ and $\epsilon_2$;
and 0.4 for the ${}^3D_3$ channel. A greater number implies heavier weight in the objective function.  

We face a high-dimensional and non-convex optimization ~\cite{carlsson2015} and therefore employ a multi-start optimization approach with 2000 initial values for the LECs randomly selected using a space-filling Latin hypercube design within conservative intervals for the LEC values, see Tab.~\ref{tab:lec_domain}. For each starting point, we find an optimum using a constrained Sequential Least Squares Programming algorithm with approximate derivatives~\cite{kraft1988,lawson1995}. The optimization also incorporates constraints that allow for up to 1\% charge symmetry breaking (CSB) or charge independence breaking in the leading-order LECs 
$\widetilde{C}_{^1S_0}^{pn}$, 
$\widetilde{C}_{^1S_0}^{nn}$, $\widetilde{C}_{^1S_0}^{pp}$.  The latter LEC is tuned separately after the optimization to empirical $pp$ effective range parameters, including Coulomb effects.

We retain the 20 candidate interactions with the lowest objective values, all of which provide an excellent description of all calibration data (see Fig.~\ref{Egs_validation} of the Supplemental Material). Finally, we perform a validation step using \emph{ab initio} many-body calculations for the ground-state energies of $^{22,24}$O and $^{40,48}$Ca. The interaction providing the best overall agreement with these validation observables is denoted \texas.

Figure~\ref{phases} shows the $np$ phase shifts computed with the \texas compared with the Granada partial wave analysis \cite{perez2013}. We also display the results from the 1.8/2.0(EM)~\cite{hebeler2011,simonis2017}, $\Delta$NNLO$_{\rm GO}$~\cite{jiang2020} and N$^3$LO$_{\rm LEFT}$~\cite{elhatisari2024} interactions. Note that the 2N sector of 1.8/2.0(EM) interaction produces the same phase shifts as the N$^3$LO$_{\rm EM}$~\cite{entem2003} 2N interaction because it is obtained by SRG evolution that leaves 2N observables invariant. As shown in Fig.~\ref{phases}, the N$^3$LO$_{\rm EM}$ gives the best description of phase shifts, but this interaction systematically underestimates charge radii. The \texas and N$^3$LO$_{\rm LEFT}$ interactions exhibit a similar level of accuracy in describing the phase shifts and can also describe charge radii well.

Table~\ref{tab:observables} lists the observables used during the optimization and validation steps as well as selected predictions.


\clearpage
\input{supplement_revised_2nd}

\end{document}

%% file: supplement_revised_2nd.tex
\onecolumngrid
\setcounter{page}{1}
\setcounter{equation}{0}
\setcounter{figure}{0}
\setcounter{table}{0}
\renewcommand{\theequation}{S\arabic{equation}}
\renewcommand{\thefigure}{S\arabic{figure}}
\renewcommand{\thetable}{S\arabic{table}}

\section*{Supplemental Material} 
\subsection{The dripline in fluorine isotopes}

\begin{figure}[!t]
\setlength{\abovecaptionskip}{0pt}
\setlength{\belowcaptionskip}{0pt}
\includegraphics[scale=0.41]{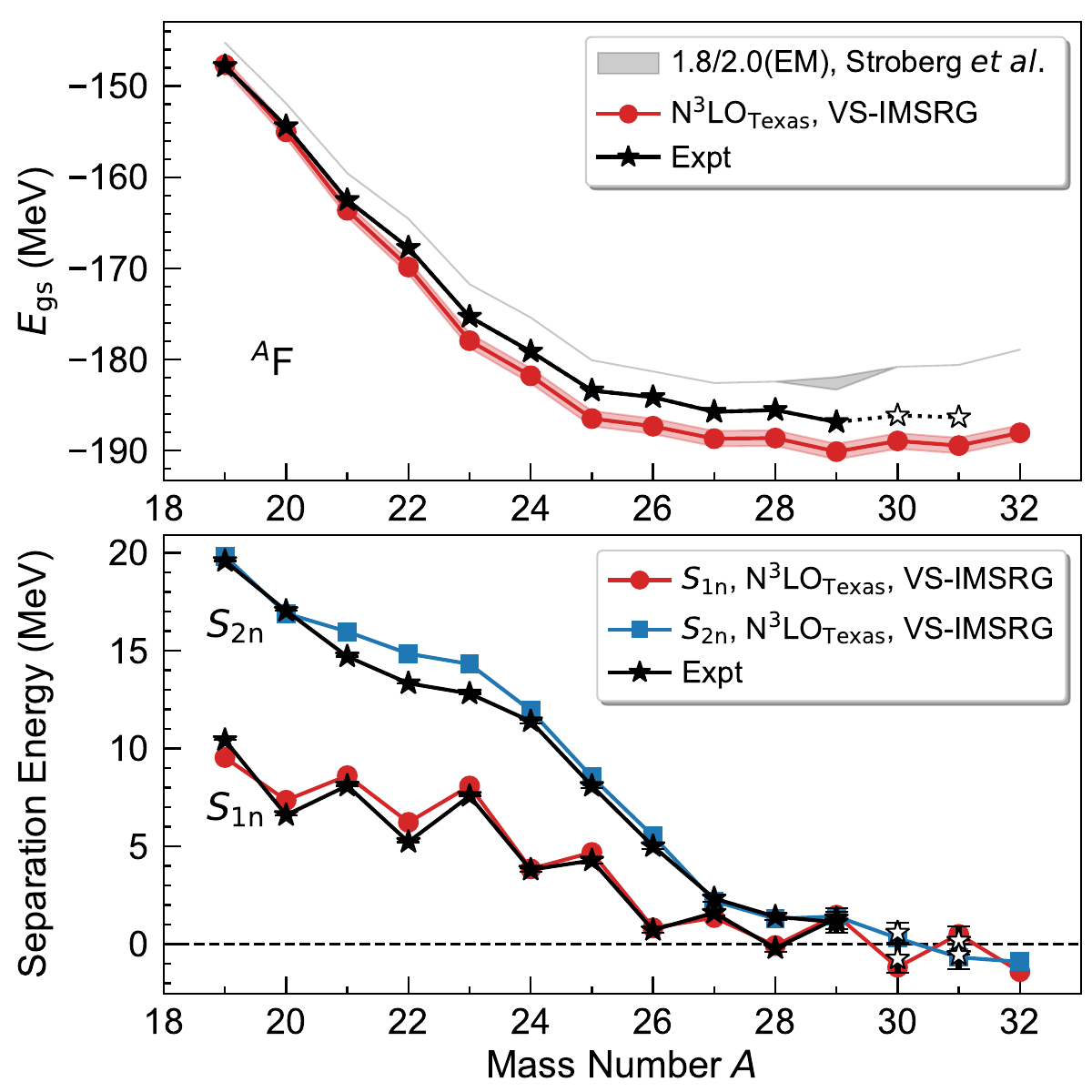} \caption{\label{F_BE_sn} 
Ground-state energies ($E_{\rm gs}$), one-neutron ($S_{1n}$) and two-neutron ($S_{2n}$) separation energies for fluorine isotopes computed with the interaction \texas and the valence-space in-medium similarity renormalization group (VS-IMSRG) method, compared with 1.8/2.0(EM) results~\cite{stroberg2021} and data \cite{ensdf,wang2021}. Stars (open) are masses (extrapolations) from AME2020 \cite{wang2021}. The red and gray bands indicate many-body uncertainty and different decoupled valence spaces, respectively.} 
\end{figure}

Figure~\ref{F_BE_sn} shows that the \texas interaction accurately reproduces the dripline in fluorine isotopes. The top panel shows ground-state energies while the bottom panel shows results for the $S_{1n}$ and $S_{2n}$. The thin red band for VS-IMSRG with \texas indicates the estimated many-body uncertainty, quantified as the $E_{\rm gs}$ difference between ``IMSRG(3f2)+Triple" and ``IMSRG(2)."  
The gray band for the 1.8/2.0(EM) shows the uncertainty given in Ref.~\cite{stroberg2021} arising from different decoupled valence spaces. 

\subsection{Dripline sensitivity to the oscillator spacing $\hbar\omega$}

\begin{figure}[!t]
\centering
\setlength{\abovecaptionskip}{0pt}
\setlength{\belowcaptionskip}{0pt}
\includegraphics[scale=0.41]{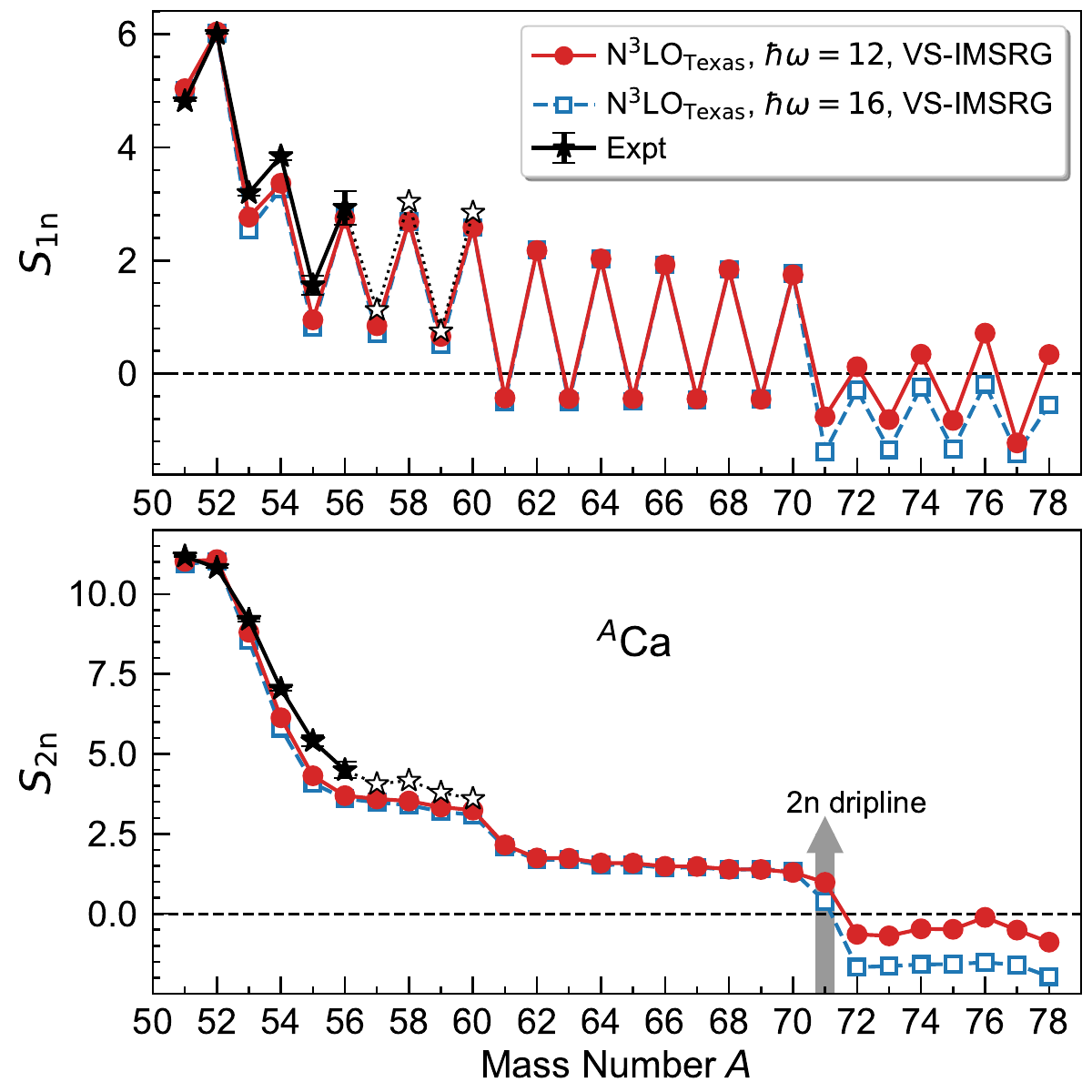} \caption{\label{Ca_Sn_hw} 
One-neutron ($S_{1n}$) and two-neutron ($S_{2n}$) separation energies for calcium isotopes computed with different $\hbar\omega$.} 
\end{figure}

Figure~\ref{Ca_Sn_hw} shows the predicted dripline in calcium isotopes for two different values of $\hbar\omega$. The $2n$ dripline is robustly placed at $^{70}$Ca for both values. 

\subsection{Excited states in $^{50,52,54,56,58}$Ca}

\begin{figure*}
\setlength{\abovecaptionskip}{0pt}
\setlength{\belowcaptionskip}{0pt}
\begin{subfigure}{}
\includegraphics[scale=0.46]{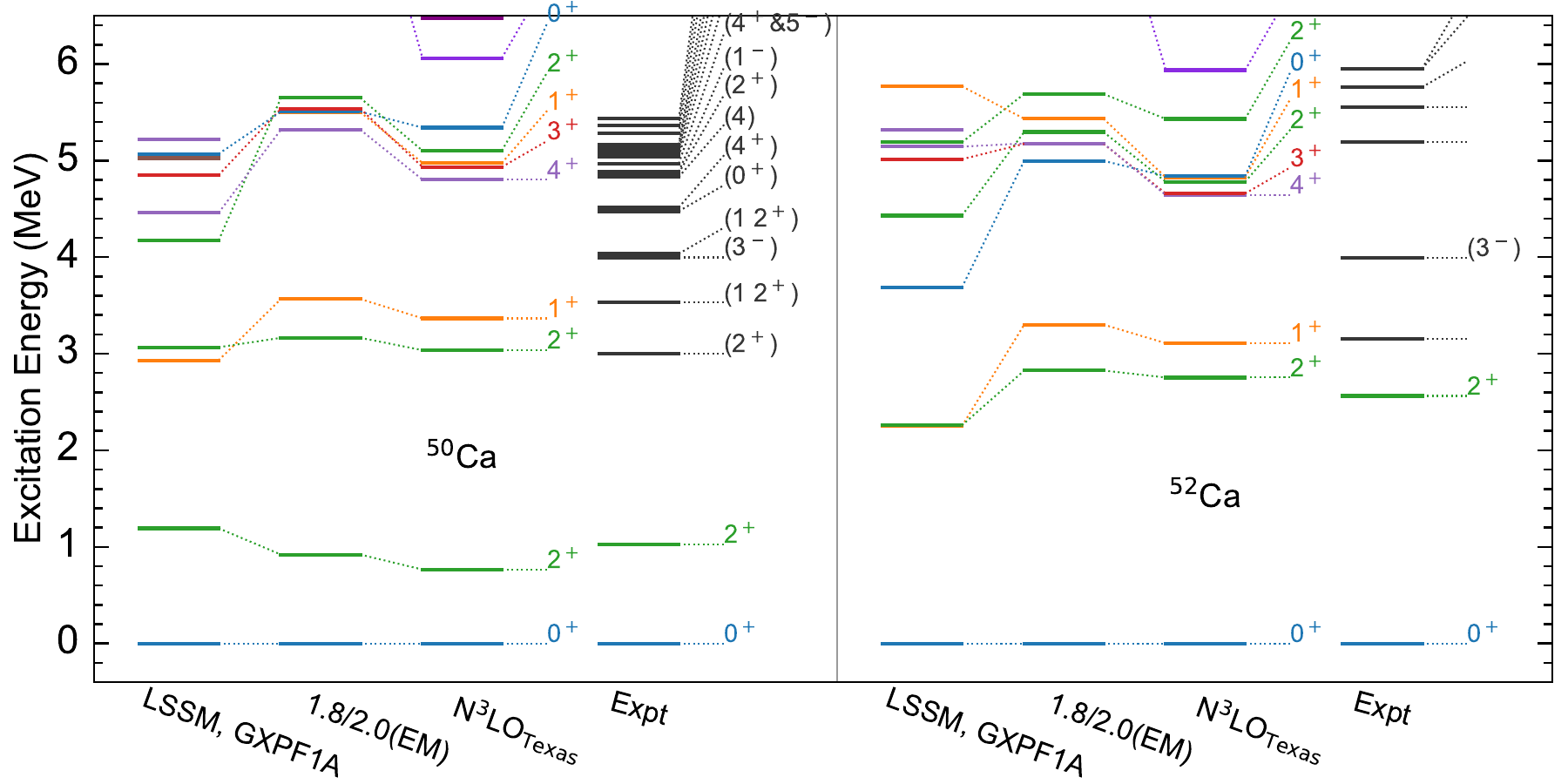}
\end{subfigure}
\\
\begin{subfigure}{}
\includegraphics[scale=0.46]{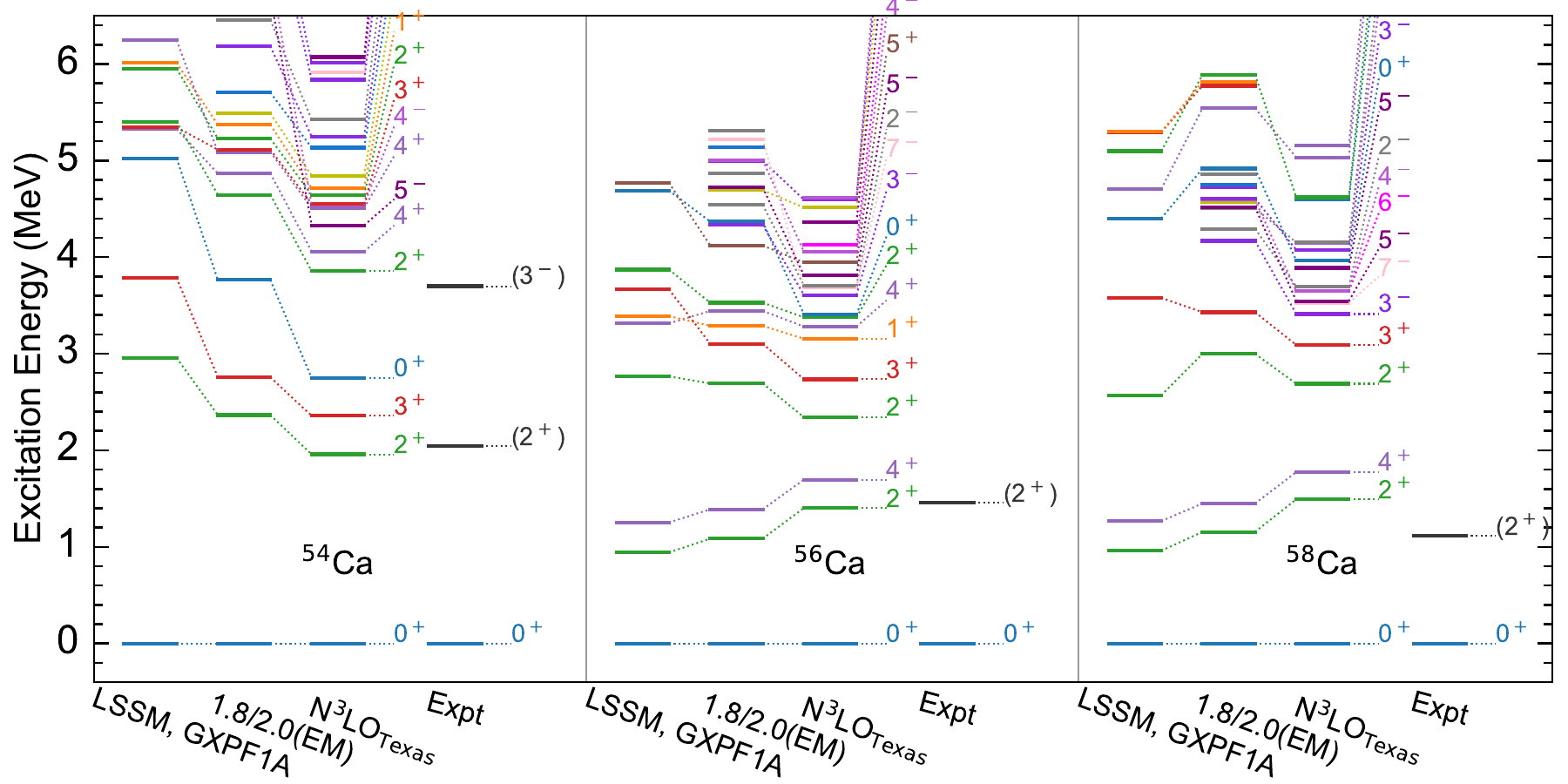}
\end{subfigure}
\caption{\label{Spectra} Excitation energies of $^{50,52,54,56,58}$Ca computed by VS-IMSRG and large-scale shell model (LSSM) with GXPF1A interaction~\cite{honma_2005}. Experimental data are from Ref.~\cite{ensdf}.}
\end{figure*}

The \texas interaction not only accurately reproduces bulk properties of the calcium isotopes but also yields realistic spectra. Fig.~\ref{Spectra} shows the low-lying excitation energies of $^{50,52,54,56,58}$Ca computed with the VS-IMSRG. As in the main text, we decouple the multishell valence-space Hamiltonian within the $^{48}$Ca core and a neutron $1p_{3/2},1p_{1/2},0f_{5/2},0g_{9/2},1d_{5/2},2s_{1/2}$ valence space from an initial model space within $e_{\rm max}$=14 and $\hbar\omega$=12 MeV. We see that \texas achieves accuracy comparable to widely used 1.8/2.0(EM) interaction for these low-lying states.

\subsection{Nuclear matter}

\begin{figure}[!t]
\centering
\setlength{\abovecaptionskip}{0pt}
\setlength{\belowcaptionskip}{0pt}
\includegraphics[scale=0.41]{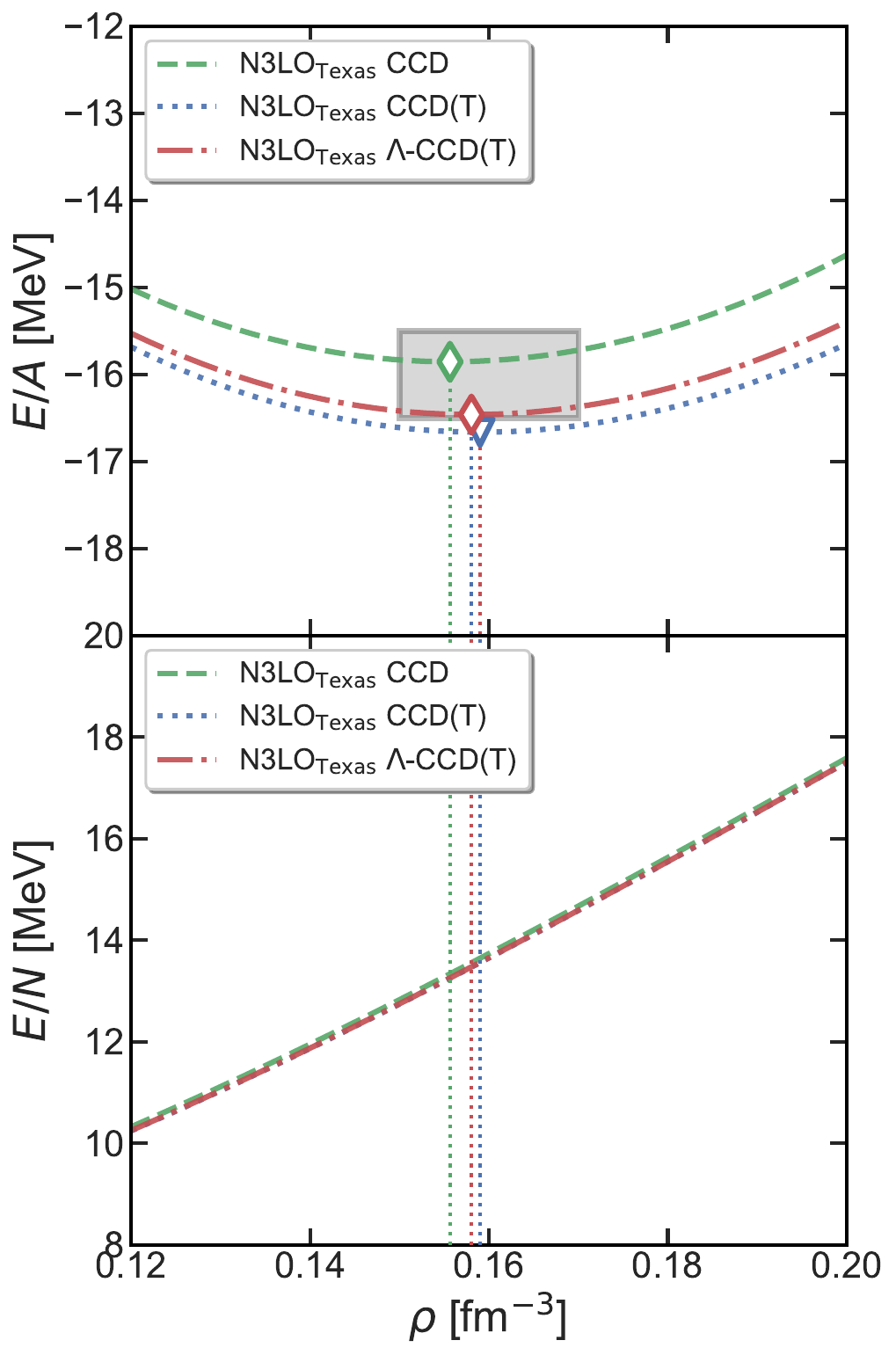} \caption{\label{nuclear_matter_} 
Energy per nucleon of symmetric nuclear matter (top panel) and pure neutron matter (bottom panel) calculated using different coupled-cluster approximations.}   
\end{figure}

Figure~\ref{nuclear_matter_} presents the comparison of different coupled-cluster approximations for nuclear matter. The results reported in the present work, shown by the red curves, are obtained with the more accurate $\Lambda$-CCD(T) method~\cite{taube2008}, which provides an improved treatment of triple correlations relative to CCD(T)~\cite{raghavachari1989}. The figure is included to clarify the quantitative differences among the various coupled-cluster approximations for both symmetric nuclear matter and pure neutron matter.

\subsection{Validation step}

\begin{figure}[!t]
\centering
\setlength{\abovecaptionskip}{0pt}
\setlength{\belowcaptionskip}{0pt}
\includegraphics[scale=0.41]{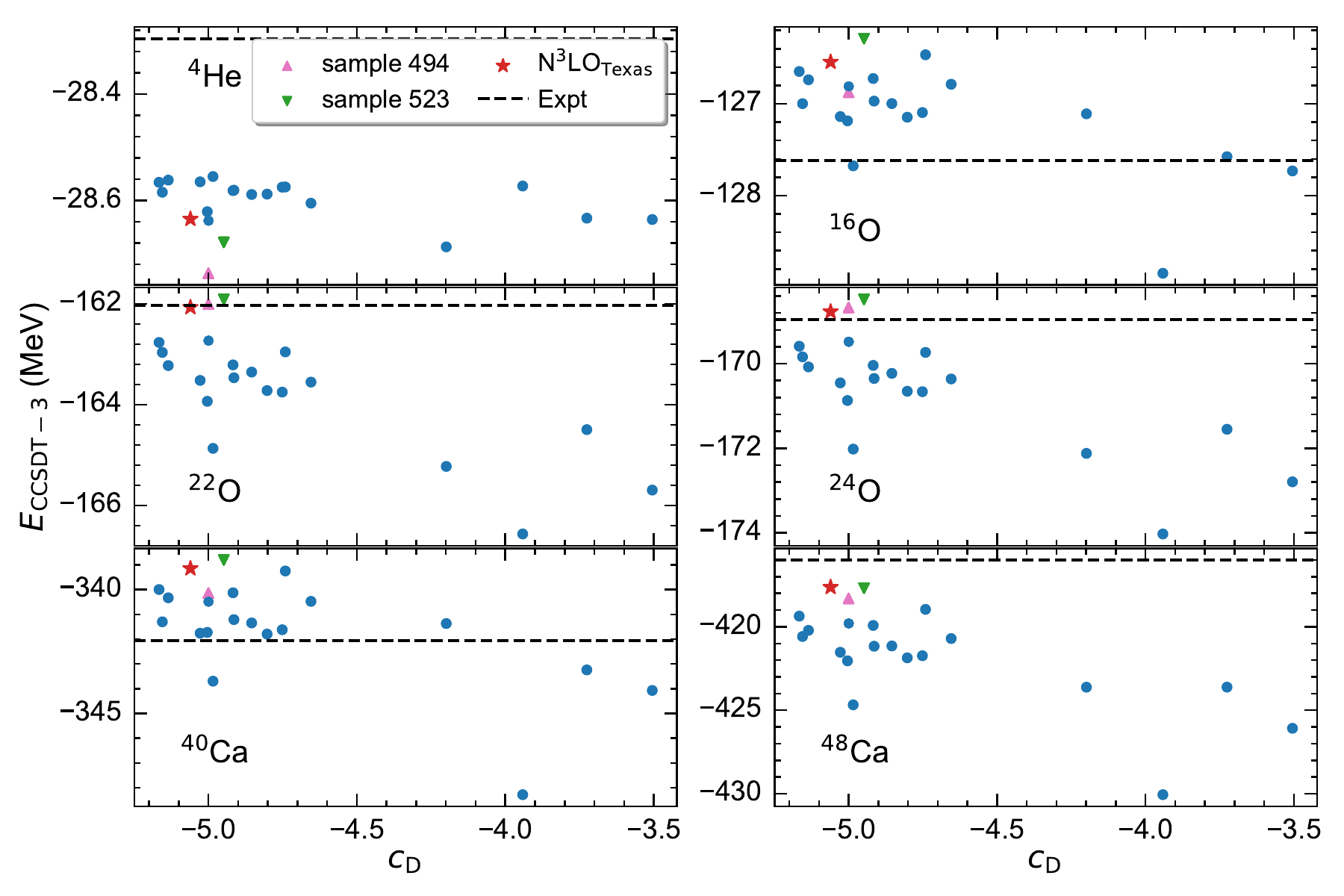} \caption{\label{Egs_validation} 
Ground-state energies of $^{4}$He, $^{16,22,24}$O and $^{40,48}$Ca computed by CCSDT-3 with 20 candidate interactions. The top three performers are \texas, and samples number 494 and 523.}   
\end{figure}

Figure~\ref{Egs_validation} shows ground-state energies for various nuclei computed with the CCSDT-3 method and the 20 candidate interactions that survived the first optimization sweep. We highlight the top three performers. 

\subsection{Cross validation of emulators used in this work}

We constructed a fast and accurate emulator for scattering phase shifts by combining Newton's variational method with ideas from eigenvector continuation \cite{melendez2021}. The absolute residuals relative to the exact solutions are less than $10^{-8}$ degrees across all partial waves, demonstrating excellent agreement with exact solutions. Using these emulators, we also computed the scattering length and effective range via effective range expansion.

For the observables in $^2$H and $^4$He, we developed emulators using the eigenvector continuation framework as presented in \cite{konig2020}, and employed 45 and 50 eigenvectors, respectively, obtained from large-space full configuration interaction calculations in a Jacobi basis. As expected, these model-driven emulators are very accurate and precise. Using 200 randomly sampled sets of LEC values, we find at most a 1\% absolute discrepancy with respect to exact calculations of the $^4$He ground-state energy and radius in the relevant ranges $[-40,-20]$ MeV and $[1.6,3.3]$ fm. For $^2$H the absolute discrepancies are below $10^{-6}$\%. 

\begin{figure*}[h!]
\setlength{\abovecaptionskip}{0pt}
\setlength{\belowcaptionskip}{0pt}
\includegraphics[scale=0.52]{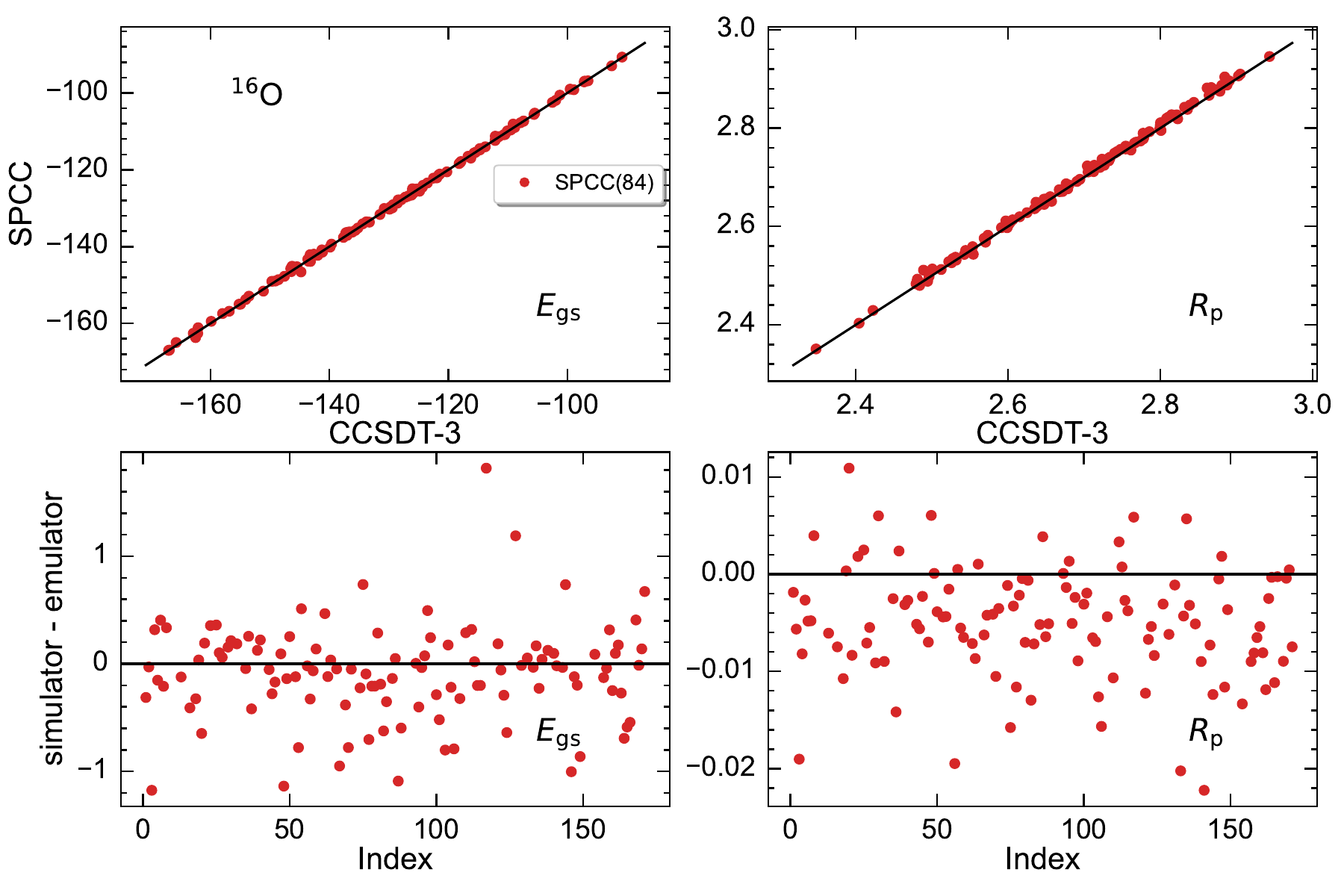}
\caption{\label{xval_o16} Cross-validation of emulator for the ground-state energy (left) and proton-point radius (right) using 200 exact CCSDT-3 calculations. Only energies between $-170$~MeV and $-90$~MeV are shown.}
\end{figure*}

We also constructed a high-precision emulator for the $^{16}$O ground-state energy and charge radius using the sub-space projected coupled-clyster (SPCC) method \cite{ekstrom2019}. To achieve higher precision in the emulator, we extended the SP-CCSD approximation used in Ref.~\cite{ekstrom2019} by including leading-order triples excitations via the CCSDT-3 method~\cite{noga1987}. Our training vectors were generated from full CCSDT-3 calculations in a model space of 13 major harmonic oscillator shells with oscillator frequency $\hbar\omega$ = 16 MeV and three-body truncation of $E_{\rm 3max}$ = 16$\hbar\omega$. We selected 84 different training points in the 28-dimensional space of LECs using a space-filling Latin hypercube design with a 20\% variation around a roughly optimized LEC set. Recognizing the large uncertainties associated with the three-body $c_{\rm D}$ and $c_{\rm E}$, we generated samples within $c_{\rm D}\in [-5, 5]$ and $c_{\rm E} \in [-3, 3]$. For each training point, we performed a full CCSDT-3 calculation to obtain the training vectors, for which we then constructed the sub-space projected norm and Hamiltonian matrices. Once these matrices are constructed, we can obtain the ground-state energy and charge radii for any target values of the LECs by diagonalizing a 84$\times$84 generalized eigenvalue problem. We quantified the accuracy of the emulator by cross-validation against 200 exact CCSDT-3 calculations of the ground-state energy and point-proton radius, and found the relative error is smaller than 1\% in both observables, as shown also in Fig.~\ref{xval_o16}. We also checked that the SPCC emulator extrapolates well outside the training region, and at the \texas point it deviates by 0.7~MeV in binding energy and 0.005~fm for the charge radius.

\begin{table}[h]
\centering
\caption{Low-energy constants (LECs) for the N$^3$LO$_{\rm Texas}$ interaction. The constants $c_i$, $\bar{d}_i$, $\widetilde{C}_i$, $C_i$, and $D_i$ (also $\widehat{D}_i$) are in units of GeV$^{-1}$, GeV$^{-2}$, $10^4$ GeV$^{-2}$, $10^4$ GeV$^{-4}$, and $10^4$ GeV$^{-6}$, respectively. The 3N LECs $c_D$ and $c_E$ are dimensionless.}
\label{tab:lecs}
\renewcommand{\arraystretch}{1.2}
\begin{tabular}{lr @{\hskip 3cm} lr}
\hline \hline
LEC & Value & LEC & Value \\
\hline
\multicolumn{4}{c}{\textbf{$\pi$N}} \\
$c_1$ & $-1.07$ & $c_2$ & $3.20$ \\
$c_3$ & $-5.32$ & $c_4$ & $3.56$ \\
$\bar{d}_1 + \bar{d_2}$ & $1.04$ & $\bar{d}_3$ & $-0.48$ \\
$\bar{d}_5$ & $0.14$ & $\bar{d}_{14} - \bar{d}_{15}$ & $-1.90$ \\
\hline
\multicolumn{4}{c}{\textbf{2N (LO)}} \\
$\widetilde{C}_{^1S_0}^{pp}$ & $-0.154291$ & 
$\widetilde{C}_{^1S_0}^{np}$ & $-0.155631$ \\ 
$\widetilde{C}_{^1S_0}^{nn}$ & $-0.155010$ &
$\widetilde{C}_{^3S_1}$ & $-0.202464$ \\
\multicolumn{4}{c}{} \\
\multicolumn{4}{c}{\textbf{2N (NLO)}} \\
$C_{^1S_0}$ & $2.514273$ &
$C_{^3P_0}$ & $1.025413$ \\
$C_{^1P_1}$ & $0.082796$ &
$C_{^3P_1}$ & $-0.977112$ \\
$C_{^3S_1}$ & $1.075598$ &
$C_{^3S_1-^3D_1}$ & $0.528490$ \\
$C_{^3P_2}$ & $-0.960907$ & & \\
\multicolumn{4}{c}{} \\
\multicolumn{4}{c}{\textbf{2N (N$^3$LO)}} \\
$\widehat{D}_{^1S_0}$ & $-1.643852$ & 
$D_{^1S_0}$ & $-19.286790$ \\
$D_{^3P_0}$ & $5.607692$  &
$D_{^1P_1}$ & $10.529344$ \\
$D_{^3P_1}$ & $5.427348$ &
$\widehat{D}_{^3S_1}$ & $5.715113$ \\
$D_{^3S_1}$ & $-40.129215$ &
$D_{^3D_1}$ & $-4.622881$ \\
$\widehat{D}_{^3S_1-^3D_1}$ & $2.852185$ &
$D_{^3S_1-^3D_1}$ & $0.505938$ \\
$D_{^1D_2}$ & $-2.369999$ & 
$D_{^3D_2}$ & $-4.984577$ \\
$D_{^3P_2}$ & $6.7635540$ &
$D_{^3P_2-^3F_2}$ & $0.188085$ \\
$D_{^3D_3}$ & $-1.196276$ & & \\
\hline
\multicolumn{4}{c}{\textbf{3N (N$^2$LO)}} \\
$c_D$             & $-5.060563$  & $c_E$             & $-1.034629$ \\
\hline
\hline
\end{tabular}
\end{table}

\begin{table}[htbp]
\centering
\caption{Initial search domains for the LEC values, where $\widetilde{C}_i$, $C_i$, and $D_i$ (also $\widehat{D}_i$) are in units of $10^4$ GeV$^{-2}$, $10^4$ GeV$^{-4}$, and $10^4$ GeV$^{-6}$, respectively. The 3N LECs $c_D$ and $c_E$ are dimensionless.}
\label{tab:lec_domain}
\begin{tabular}{lrr @{\hskip 3cm} lrr}
\toprule
LEC & Min & Max & LEC & Min & Max \\
\hline
$\widetilde{C}_{^1S_0}^{pp}$     &  $-0.30$ &  $-0.10$ &
$\widetilde{C}_{^1S_0}^{np}$      &  $-0.30$ &  $-0.10$ \\
$\widetilde{C}_{^1S_0}^{nn}$      &  $-0.30$ &  $-0.10$ &
$\widetilde{C}_{^3S_1}$        &  $-0.30$ &  $-0.10$ \\
$C_{^1S_0}$         &   $2.30$ &   $2.90$ &
$C_{^3P_0}$         &   $0.80$ &   $1.40$ \\
$C_{^1P_1}$         &  $-1.00$ &   $1.00$ &
$C_{^3P_1}$         &  $-1.40$ &   $1.50$ \\
$C_{^3S_1}$         &   $0.10$ &   $1.70$ &
$C_{^3S_1-^3D_1}$     &   $0.00$ &   $1.20$ \\
$C_{^3P_2}$         &  $-2.10$ &   $2.40$ &
$\widehat{D}_{^1S_0}$        &  $-2.20$ &   $4.50$ \\
$D_{^1S_0}$         & $-28.00$ & $-18.00$ &
$D_{^3P_0}$         &   $3.90$ &   $6.40$ \\
$D_{^1P_1}$         &   $8.50$ &  $18.00$ &
$D_{^3P_1}$         &   $2.00$ &  $10.50$ \\
$\widehat{D}_{^3S_1}$        &  $-9.70$ &   $9.30$ &
$D_{^3S_1}$         & $-40.60$ &  $16.70$ \\
$D_{^3D_1}$         &  $-8.20$ &   $4.50$ &
$\widehat{D}_{^3S_1-^3D_1}$    & $-10.00$ &   $9.30$ \\
$D_{^3S_1-^3D_1}$     &  $-7.40$ &  $10.00$ &
$D_{^1D_2}$         &  $-2.60$ &   $1.10$ \\
$D_{^3D_2}$         &  $-8.00$ &  $-2.00$ &
$D_{^3P_2}$         &  $-3.00$ &  $15.30$ \\
$D_{^3P_2-^3F_2}$    &  $-2.60$ &   $2.90$ &
$D_{^3D_3}$         &  $-3.00$ &   $1.00$ \\
$c_D$             &  $-8.00$ &   $8.00$ &
$c_E$              &  $-8.00$ &   $8.00$ \\ 
\hline \hline
\end{tabular}
\end{table}

\twocolumngrid